\documentclass[a4paper,twoside]{article}

\usepackage{epsfig}
\usepackage{subcaption}
\usepackage{calc}
\usepackage{amssymb}
\usepackage{amstext}
\usepackage{amsmath}
\usepackage{amsthm}
\usepackage{multicol}
\usepackage{pslatex}
\usepackage{apalike}
\usepackage{hyperref}
\usepackage{xurl}
\usepackage{algorithm2e}
\usepackage[bottom]{footmisc}
\usepackage{makecell}
\usepackage{multirow}
\usepackage{SCITEPRESS}    

\begin{document}

\title{Adversarial Co-Evolution of Malware and Detection Models: A Bilevel Optimization Perspective}

\author{\authorname{Olha Jurečková\sup{1}\orcidAuthor{0000-0002-8858-4826},  Martin Jureček\sup{1}\orcidAuthor{0000-0002-6546-8953}, Matou\v{s} Koz\'{a}k\sup{1}\orcidAuthor{0000-0001-8329-7572} and R\'{o}bert L\'{o}rencz\sup{1}\orcidAuthor{0000-0001-5444-8511}}
\affiliation{\sup{1}Faculty of Information Technology, Czech Technical University in Prague, \mbox{Thákurova 9}, Prague, 16000, Czech Republic}
\email{\{jurecolh, martin.jurecek, kozakmat, robert.lorencz\}@fit.cvut.cz}}


\abstract{
Machine learning-based malware detectors are increasingly vulnerable to adversarial examples. Traditional defenses, such as one-shot adversarial training, often fail against adaptive attackers who use reinforcement learning to bypass detection. This paper proposes a robust defense framework based on bilevel optimization, explicitly modeling the strategic interaction between a defender and an attacker as an adversarial co-evolutionary process. We evaluate our approach using the MAB-malware framework against three distinct malware families: Mokes, Strab, and DCRat. Our experimental results demonstrate that while standard classifiers and basic adversarial retraining often remain vulnerable, showing evasion rates as high as 90\%, the proposed bilevel optimization approach consistently achieves near-total immunity, reducing evasion rates to 0--1.89\%. Furthermore, the iterative framework significantly increases the attacker's query complexity, raising the average cost of successful evasion by up to two orders of magnitude. These findings suggest that modeling the iterative cycle of attack and defense through bilevel optimization is essential for developing resilient malware detection systems capable of withstanding evolving adversarial threats.
}

\onecolumn \maketitle \normalsize \setcounter{footnote}{0} \vfill

\section{\uppercase{Introduction}}
\label{sec:introduction}

The number of malicious files detected worldwide continues to grow rapidly, with Kaspersky reporting an average of 500,000 detections per day in 2025, representing a 7\% increase compared to the previous year \cite{kaspersky2025malware}. This pattern demonstrates the growing complexity of contemporary malware as well as the urgent need for cybersecurity approaches that can effectively adapt to rapidly evolving threats.
 
Malware detection faces a fundamental security challenge arising from the asymmetry between adaptive attackers and learning-based defenders. While defenders rely on detection models trained on previously observed data, attackers can rapidly generate new malware variants specifically designed to evade these defenses. This problem is reflected in the literature, which reveals that no single detection approach can identify all new-generation and sophisticated malware \cite{aslan2020comprehensive}. 

In practice, attackers continuously adapt their strategies, often matching or surpassing advances in antivirus (AV) detection systems. Malware classification thus operates in a fundamentally adversarial environment, where binaries are deliberately modified to evade detection using techniques such as polymorphism, packing, and other obfuscation methods. While adversarial machine learning explores evasion strategies including reinforcement learning and perturbation-based attacks \cite{kozak2024creating,aryal2024survey,ban2024empirical}, these works largely focus on attacking fixed classifiers and do not capture the strategic, sequential interaction between attackers and defenders. 

Malware detection remains an open challenge due to persistent distribution shifts caused by evolving malware and new benign software \cite{kaspersky_ml_malware_detection}. Unlike static machine learning settings, this environment requires continuous data collection and periodic retraining. However, the inherent latency and costs associated with retraining create exploitable windows where outdated models become vulnerable, turning distribution shifts into a fundamental security risk.

The primary contribution of this work lies in the formulation of malware detection as an iterative, co-evolutionary game between an adaptive attacker and a defender, modeled through a bilevel optimization perspective. Unlike traditional static defenses, we systematically explore the convergence properties and long-term behavior of this interaction by employing an iterative framework where the defender's model and the attacker's strategies evolve in response to each other. We demonstrate that this strategic co-adaptation, integrating adversarial samples from successive attack rounds back into the training process, not only identifies the fundamental limitations of standard one-shot adversarial training but also leads to a practical equilibrium that significantly enhances detection robustness and escalates the operational cost for the attacker across diverse malware families.

The remainder of this paper is organized as follows. Section~2 reviews related work in adversarial malware generation, and game-theoretic modeling of security problems. Section~3 provides the necessary background on adversarial malware generation. Section~4 details our research methodology, explaining the iterative co-evolutionary process and the evaluation framework. Section~5 introduces our game-theoretic framework and formulates the interaction between the attacker and the defender as a bilevel optimization problem. Section~6 presents the experimental results, including the analysis of convergence and exploitability in the proposed setting. Finally, Section~7 concludes the paper and outlines directions for future work.

\section{\uppercase{Related Work}}
\label{sec:rl}

This section reviews key research on security games in malware detection, adversarial machine learning, and adversarial classification that has influenced the development of our framework.

Several studies have examined adversarial malware generation to challenge machine learning–based classifiers. In \cite{kozak2023combining}, the authors  propose combining multiple generators, with the random AMG followed by MAB-Malware combination achieving the highest evasion rate of 15.9\% against leading antivirus products. In a complementary study \cite{louthanova2024comparison}, the authors evaluate five generators, PartialDOS, FullDOS, GAMMA padding, GAMMA section-injection, and Gym-malware, demonstrating that the reinforcement learning–based Gym-malware generator generates samples most efficiently, averaging 5.73 seconds per sample. 

The authors of \cite{aryal2024survey} provide a comprehensive survey of adversarial attacks and defenses in malware analysis, highlighting the limitations of current approaches and the vulnerability of machine learning detectors. Similarly, Suciu et al. \cite{suciu2019exploring} show that adversarial attacks are less effective on models trained with large datasets, while architectures like MalConv, which lack positional feature encoding, remain particularly susceptible to append-based attacks, offering insights into the generalization and practical effectiveness of adversarial malware examples. Together, these works underscore the challenges of static classifiers and single-strategy attacks, motivating defense mechanisms such as our security game framework, which anticipates adaptive, strategic attackers and accounts for coordinated, multi-family evasion strategies.

Stackelberg Security Games (SSGs) offer a principled framework for modeling such strategic interactions. Wilczyński et al. \cite{wilczynski2016stackelberg} provide a comprehensive overview of SSG models, applications in physical and cyber security, and computational methods for finding optimal strategies. In \cite{sinha2018stackelberg}, the authors review over a decade of SSG research, emphasizing algorithmic advancements, modeling improvements, and practical deployments, while highlighting challenges such as uncertainty, resource constraints, and dynamic attacker behavior. 

Complementing these works, the authors of \cite{dritsoula2017game} model adversarial classification as a nonzero-sum game, where a defender and attacker strategically interact over class 1 data while class 0 data remain random. Their analysis of Nash equilibria shows that attackers tend to mimic normal behavior, defenders should randomize classification rules, and optimal strategies often rely on a small set of threshold-based classifiers. Together, these studies provide a strong foundation for designing robust, game-theoretic defenses against adaptive and strategic malware attackers, motivating our security game framework.

The authors of \cite{zhuang2025robust} propose two realistic threat models, M-TBA and AM-TBA, to characterize black-box adversarial behaviors with different levels of adaptability under traffic-space constraints. They present NashAE, a game-theoretic ensemble malware detector that achieves robustness by approximating a Nash equilibrium and modeling the attacker–defender interaction as a minimax game. Since the minimax game has no closed-form solution and cannot be efficiently handled by typical gradient-based methods, they decompose it into two related Markov Decision Processes and solve them using Bayesian optimization to obtain an efficient and responsive solution. They further evaluate their proposed attack and defense framework on two encrypted malware traffic datasets using ten detection strategies under both M-TBA and AM-TBA settings.

In \cite{ebrahimi2022adversarial}, the authors propose a game-theoretic framework for malware detection using adversarial reinforcement learning. The interaction between the malware detector and an adaptive attacker is represented by a minimax optimization game, with adversarial malware variants developed using a reinforcement learning-based attack policy to improve robustness. The approach alternates between strengthening the attacker and improving detector resilience, and it is tested on a 33.2 GB real-world malware dataset collected from VirusTotal\footnote{\url{https://virustotal.com}}.

\section{\uppercase{Background}}
\label{sec:back}

This section establishes the technical foundations of our study. We first formalize adversarial attacks on Portable Executable (PE) files\footnote{\url{https://learn.microsoft.com/en-us/windows/win32/debug/pe-format}}, emphasizing the constraints of functionality preservation. Subsequently, we describe the MAB-malware framework, a reinforcement learning-based agent used to simulate an adaptive attacker capable of discovering optimal evasion strategies.

\subsection{Adversarial Attacks on PE Files}
\label{subsec:background_pe_attacks}

The objective of an adversarial attack in the malware domain is to transform a malicious binary $x \in \mathcal{X}$ into an adversarial sample $x_{adv} = x + \delta$, where $\delta$ represents a sequence of perturbations. In the context of feature-based detection, let $\phi: \mathcal{X} \to \mathbb{R}^n$ be a mapping function (e.g., EMBER) that extracts a feature vector $z = \phi(x)$. An adversarial sample is successful if $f(\phi(x + \delta)) = \text{benign}$, while the original label is $f(\phi(x)) = \text{malware}$.

Unlike adversarial examples in continuous domains such as computer vision, perturbations in the Portable Executable (PE) space are constrained by the requirement of \textit{semantic integrity}. As emphasized in \cite{kozak2024creating}, any modification $\delta$ must result in a valid binary that remains executable and preserves its original malicious payload. Formally, this restricts the perturbation to a discrete action space $\mathcal{S}(x)$, such that:
\begin{equation}
    x_{adv} = x + \delta \quad \text{s.t.} \quad \delta \in \mathcal{S}(x) \text{ and } \text{Sem}(x) = \text{Sem}(x_{adv}),
\end{equation}
where $\text{Sem}(\cdot)$ represents the functional behavior of the program.

The fundamental purpose of these adversarial samples is to probe and exploit the decision boundaries of machine learning models. By mimicking the statistical properties of benign software, such as through section injection or import table manipulation, these samples expose the "blind spots" of a classifier, forcing it to rely on non-robust features that are easily susceptible to strategic manipulation.

\subsection{MAB-malware}

The MAB-Malware approach \cite{song2021mabmalware} leverages reinforcement learning to generate adversarial malware samples by modeling the attack process as a multi-armed bandit (MAB) problem. Its objective is to discover a small subset of binary transformations that leads a target classifier to incorrectly classify malicious inputs as benign. The procedure consists of two main stages:

\begin{enumerate}
\item[(1)] an exploration stage, where the MAB agent sequentially selects and applies candidate macro- and micro-level transformations until either evasion is achieved or a predefined modification budget is reached, and
\item[(2)] a refinement stage, during which individual transformations are systematically re-evaluated and discarded if their removal does not compromise the evasion outcome.
\end{enumerate}

Because the MAB framework assumes independence among actions (i.e., no explicit ordering or interaction constraints), this subsequent pruning step can effectively reduce the number of applied modifications while maintaining successful evasion.

Representative transformations include appending benign content (e.g., overlays or new sections), inserting or renaming sections, nullifying certificate or debug-related fields, altering optional header checksums, and applying semantics-preserving code changes. In our experiments, MAB-Malware was employed to craft adversarial samples targeting a classifier trained on features from EMBER dataset \cite{anderson2018ember}.

\section{\uppercase{Methodology}}
\label{sec:methodology}

The goal of this work is to systematically compare malware detection performance under three different approaches:
(1) a standard machine learning classifier (malware detector) trained on non-adversarial data,
(2) the same classifier subsequently retrained on adversarial samples generated using MAB-malware, and
(3) a classifier optimized via bilevel optimization, which explicitly models the interaction between the attacker and the defender.

\subsection{Data Preparation and Experiment (1)}

At the beginning, the dataset is split into three disjoint subsets: the training set $D^{train}$, the validation set $D^{val}$, and the test set $D^{test}$. All subsets contain original (non-adversarial) malware binaries as well as benign binaries.

The first experiment consists of training a standard machine learning classifier, Random Forest, exclusively on non-adversarial samples from $D^{train}$. The validation set $D^{val}$ is used for hyperparameter tuning of Random Forest. The final performance and robustness of the model are evaluated using accuracy, recall, and precision on $D^{test}$. During testing, the model is also exposed to samples modified using MAB-malware to establish a baseline level of robustness against adversarial attacks. The test set remains fixed across all experiments (1), (2), and (3), and is never used during training or adversarial sample generation to ensure objective evaluation.

\subsection{Adversarial Training Inclusion -- Experiment (2)}

In the second experiment, we use the trained model from Experiment (1) as the target classifier for MAB-malware. The malware samples from $D^{train}$ are modified using MAB-malware with the objective of generating adversarial samples that evade detection by the target model.

The resulting augmented dataset $D^{train}_{adv}$, which consists of both original and adversarial samples (all labeled correctly as malware), is used to retrain the classifier from Experiment (1). The resulting more robust model is then evaluated again using accuracy, recall, and precision on the unchanged test set $D^{test}$.

\subsection{Bilevel Optimization and Dynamic Game -- Experiment (3)}

The third and most complex experiment employs bilevel optimization to simulate a strategic interaction between the defender and the attacker. Careful data handling is essential to preserve evaluation integrity.

\textbf{Initialization:} The defender trains an initial model on $D^{train}$ (as in Experiment (1)).

\textbf{Initial Attack:} The attacker applies MAB-malware to samples from $D^{train}$ using the initial model as the target classifier. This produces the first set of successful adversarial samples $(D^{train}_{adv})_0$ (as in Experiment (2)).

\textbf{Iterative Cycle (for $i = 1$ to $\text{Max\_iter}$):}

\begin{itemize}
    \item \textbf{Defender:} The defender retrains the classifier using the original dataset $D^{train}$ together with adversarial samples generated in previous iterations $(D^{train}_{adv})_j$, where $j \leq i-1$. All samples retain correct labels, meaning adversarial samples are still labeled as malware.
    
    \item \textbf{Attacker:} MAB-malware is applied again to the original malware samples from $D^{train}$, but now using the updated classifier as the target model. This produces a new set of adversarial samples $(D^{train}_{adv})_i$, reflecting the attacker’s adaptation to the current defense.
\end{itemize}

After completing the iterations (with $\text{Max\_iter}$ determined empirically), the final optimized model is evaluated on $D^{test}$. The key difference from previous experiments is that the target classifier evolves in each iteration, forcing the defender to learn more robust parameters that are not easily circumvented in subsequent attacker steps.

The overall bilevel optimization procedure is summarized in Algorithm~\ref{alg:bilevel}.

\begin{algorithm}
\caption{Bilevel Optimization for Adversarial Malware Generation}
\label{alg:bilevel}

\KwIn{Training dataset $D^{train}$, maximum iterations $\text{Max\_iter}$}
\KwOut{Optimized classifier $f^{*}$}

\textbf{Initialization:} \\
Train initial classifier $f_0$ on $D^{train}$ \\
$(D^{train}_{adv})_0 \leftarrow \text{MAB-malware}(f_0, D^{train})$

\For{$i = 1$ \KwTo $\text{Max\_iter}$}{

    \textbf{Defender step:} \\
    $D^{train}_{aug} \leftarrow D^{train} \cup \bigcup_{j=0}^{i-1} (D^{train}_{adv})_j$ \\
    $f_i \leftarrow \text{Train}(D^{train}_{aug})$

    \textbf{Attacker step:} \\
    $(D^{train}_{adv})_i \leftarrow \text{MAB-malware}(f_i, D^{train})$
}

\Return{$f_{\text{Max\_iter}}$}

\end{algorithm}

While the assumption that the attacker has full access to the target model $f_i$ corresponds to a white-box attack scenario, it allows us to evaluate the defender's robustness under the worst-case conditions, following Kerckhoffs's principle. This ensures that the resulting defense boundary is resilient even against highly informed adversaries.

\subsection{Final Evaluation}

In the final evaluation, we assess the detection performance of the models from all three experimental scenarios. To ensure a comprehensive assessment of both classification quality and security, each model is evaluated on two distinct datasets: the \textit{clean test set} (containing only original, non-adversarial samples) and the \textit{adversarial test set} (containing samples actively manipulated by the MAB-malware agent to evade detection).

At each step, the target classifier used by the MAB-malware agent during the evaluation phase corresponds exactly to the most recently trained defender model from the respective experiment. All experiments are conducted independently for each malware family (Mokes, Strab, and DCRat), resulting in a systematic comparison of the three defensive approaches across different adversarial behaviors. All experimental settings are evaluated using the Random Forest architecture to maintain consistency in the comparison of the defensive strategies.

\section{\uppercase{Game-Theoretic and Bilevel Formulation}} 
\label{sec:proposed}

The interaction between a defender (malware detector) and an adaptive attacker (MAB-malware generator) can be formalized through the lens of game theory, specifically as a \textbf{bilevel optimization problem}. This framework allows us to distinguish between static defenses (Experiment 1), simple reactive retraining (Experiment 2), and strategic co-evolution (Experiment 3).

\subsection{Formal Framework}

Let $f_{\theta}: \mathcal{X} \to \mathcal{Y}$ be a classifier parameterized by $\theta \in \Theta$, representing the defender's detection model. In this work, the interaction is modeled as a strategic game where the defender aims to correctly identify malware, while the role of the adaptive attacker is realized by the MAB-malware framework. The attacker seeks functionality-preserving transformations $\delta \in \mathcal{S}(x)$ to maximize the probability of evasion. This strategic interaction leads to the following bilevel formulation:
\begin{equation}
    \min_{\theta \in \Theta} \; \frac{1}{|D^{train}|} \sum_{(x,y) \in D^{train}} \left[ \mathcal{L}(f_{\theta}(x + \delta^*(\theta)), y) \right]
\end{equation}
\begin{equation}
    \text{s.t. } \delta^*(\theta) \in \arg\max_{\delta \in \mathcal{S}(x)} \mathcal{L}_{\text{atk}}(f_{\theta}(x + \delta), y),
\end{equation}
where $f_\theta$ is the classifier parameterized by $\theta$ within the parameter space $\Theta$. The formulation seeks to minimize the \textit{average loss} over all samples $(x, y)$ in the training set $D^{train}$. The functions $\mathcal{L}$ and $\mathcal{L}_{\text{atk}}$ represent the loss functions for the defender and the attacker, respectively. Specifically, $\mathcal{L}_{\text{atk}}$ is defined such that it is maximized when the classifier $f_\theta$ misclassifies a malicious sample as benign, representing a successful evasion. The set $\mathcal{S}(x)$ defines the space of functionality-preserving transformations available for a given sample $x$, from which the attacker chooses an optimal perturbation $\delta^*$.

The inner problem represents the attacker's best response $\delta^*(\theta)$ to a fixed model $\theta$, while the outer problem represents the defender's quest for parameters $\theta$ that are robust to that optimal response.

\subsection{Modeling the Three Experimental Scenarios}

The methodology described in Section~\ref{sec:methodology} maps directly to different approximations of this game-theoretic equilibrium:

\begin{enumerate}
    \item \textbf{Baseline (Exp 1):} This represents a \textit{non-strategic} defender. The defender assumes $\delta = 0$, solving only the outer minimization on $D^{train}$ without considering the inner maximization.
    
    \item \textbf{Adversarial Retraining (Exp 2):} This is a \textit{one-step reactive} defense. The defender first solves the inner maximization against the baseline model $f_{\theta_0}$ to generate $(D^{train}_{adv})_0$, then performs a single update to $\theta$ by training on the augmented set $D^{train} \cup (D^{train}_{adv})_0$.
    
    \item \textbf{Bilevel IBR (Exp 3):} This represents the \textit{strategic} defender. It approximates the full bilevel solution through \textbf{Iterative Best-Response (IBR)} dynamics.
\end{enumerate}

\subsection{Iterative Best-Response Dynamics}

As detailed in Algorithm~\ref{alg:bilevel}, the solution to the bilevel problem is approximated through a sequence of strategic interactions over $i = 1, \dots, \text{Max\_iter}$. At each iteration $i$, the dynamics are defined as follows:

\begin{itemize}
    \item \textbf{Attacker Step:} The attacker utilizes the MAB-malware framework to solve the inner maximization problem against the current defender's model $f_i$:
    \begin{equation}
        (D^{train}_{adv})_i = \{ x + \delta \mid \delta \approx \arg\max_{\delta \in \mathcal{S}(x)} \mathcal{L}_{\text{atk}}(f_i(x+\delta), y) \}.
    \end{equation}
    
    \item \textbf{Defender Step:} The defender updates the model by minimizing the average loss over the cumulative history of all preceding attacks:
    \begin{equation}
        f_{i+1} = \text{Train}\left( D^{train} \cup \bigcup_{j=0}^{i} (D^{train}_{adv})_j \right).
    \end{equation}
\end{itemize}

This iterative cycle aims to reach a practical \textbf{$\epsilon$-Nash Equilibrium}, where the defender's model $f^*$ is optimized against the most effective adversarial samples, and the attacker's success is minimized.  The equilibrium is reached when the iterative updates to $f_i$ no longer yield a significant reduction in the evasion rate, effectively terminating the co-evolutionary process. Given the discrete nature of the PE file transformation space $\mathcal{S}(x)$, we monitor convergence through the \textit{stability of the evasion rate} across successive iterations.

\section{\uppercase{Experimental Results}}
\label{sec:experiments}

This section details our experimental findings. First, we describe the dataset and define the metrics for classification and robustness. Next, we present and analyze the empirical results for the Mokes, Strab, and DCRat families across all three defensive strategies. Finally, we provide a comparative discussion highlighting the effectiveness of the bilevel optimization framework in neutralizing adaptive adversarial attacks.

\subsection{Dataset}

In the experimental evaluation, we consider three malware families: Mokes, DCRat, and Strab. These families were selected because MAB-malware with the target classifier EMBER achieved high success rates on them \cite{jureckova2026detecting}. Specifically, for Mokes we generated 2058 adversarial binaries out of 2216, for DCRat 1010 out of 1026, and for Strab 1595 out of 2191.

The feature set employed in our experiments is based on the EMBER 2018 dataset, as described in \cite{anderson2018ember}. Malware binaries for the Mokes, Strab, and DCRat families are sourced from the RawMal-TF dataset \cite{balik2025rawmal}. In this dataset, each malware family is accompanied by its corresponding feature vectors and a balanced set of feature vectors representing benign files, which were derived from the original EMBER dataset.

For our experimental evaluation, we do not use the full dataset. Instead, all reported results are obtained on a balanced subset of 500 samples per malware family, composed in a 50:50 ratio of malware and benign feature vectors (i.e., 250 malicious and 250 benign samples). This subset-based evaluation enables a consistent comparison across the three defensive strategies while keeping the iterative bilevel optimization experiments computationally tractable.

\subsection{Evaluation Metrics}
\label{subsec:metrics}

We evaluate classification performance using Accuracy ($Acc$), Precision ($Prec$), and Recall ($Rec$), defined as:
\begin{align}
Acc &= \frac{TP+TN}{TP+TN+FP+FN}, \nonumber \\
Prec &= \frac{TP}{TP+FP}, \quad Rec = \frac{TP}{TP+FN},
\end{align}
where $TP$, $TN$, $FP$, $FN$ denote true/false positives and negatives.

Adversarial robustness is measured by:
\begin{itemize}
    \item \textbf{Evasion Rate (ER):} The percentage of malware samples that successfully bypass detection: $ER = (M_{\text{evasive}} / M_{\text{total}}) \times 100\%$, where $M_{evasive}$ is the count of successful adversarial samples, and $M_{total}$ is a total number of files submitted to that classifier.
    \item \textbf{Average Queries (Avg Q):} The mean number of queries required by the MAB-malware agent to find a successful perturbation. This serves as a proxy for \textit{attack complexity}.
\end{itemize}
All metrics are reported on the disjoint test set $D^{test}$, considered in two variants: a \textit{clean} version containing only the original (non-adversarial) samples, and a \textit{clean + adversarial} version in which the clean test set is extended with MAB-malware-perturbed counterparts of its malicious samples, while the benign samples remain unchanged. In the remainder of the paper, we refer to the latter as the \textit{adversarial test set} for brevity.

\begin{table*}[h]
\centering
\caption{Comprehensive Performance for Mokes (Clean vs. Adversarial Test Sets)}
\label{tab:combined_mokes}
\begin{tabular}{l|ccc|ccccc}
\hline
& \multicolumn{3}{c|}{\textbf{Clean Test Set}} & \multicolumn{5}{c}{\textbf{Adversarial Test Set}} \\
\textbf{Experiment} & \textbf{Acc} & \textbf{Prec} & \textbf{Rec} & \textbf{Acc} & \textbf{Prec} & \textbf{Rec} & \textbf{ER (\%)} & \textbf{\makecell{Avg\\Q}} \\ \hline
(1) Baseline    & 1.00 & 1.00 & 1.00 & 0.68 & 1.00 & 0.51 & 95.83 & 15.93 \\
(2) Adv. Train  & 1.00 & 1.00 & 1.00 & 0.97 & 1.00 & 0.94 & 6.25  & 932.33 \\
(3) Bilevel     & 1.00 & 1.00 & 1.00 & \textbf{1.00} & \textbf{1.00} & \textbf{1.00} & \textbf{0.00} & -- \\ \hline
\end{tabular}
\end{table*}

\subsection{Evaluation of the Mokes Malware Family}

The experimental results, summarized in Table \ref{tab:combined_mokes}, demonstrate a significant evolution in model robustness across the three tested approaches. In the baseline scenario (Experiment 1), the classifier achieved perfect scores but proved to be highly vulnerable to adversarial manipulation, with an evasion rate (ER) of 95.83\%. The attacker required only 15.93 queries on average to bypass the detector.

The inclusion of adversarial training in Experiment 2 substantially improved the defense, dropping the ER to 6.25\%. However, as shown in Table \ref{tab:combined_mokes}, the bilevel optimization (Experiment 3) provided the most complete defense. When evaluated against the adversarial test set, which includes actively manipulated samples, the bilevel model maintained perfect precision, recall, and a 0\% evasion rate, effectively neutralizing the attack.

\subsection{Evaluation of the Strab Malware Family}

The Strab family reveals distinct characteristics, as shown in Table \ref{tab:combined_strab}. The baseline classifier achieved an accuracy of 0.90 on clean (i.e., non-adversarial) samples but suffered an ER of 77.36\%. In Experiment 2, retraining provided only moderate improvements, leaving the model vulnerable with a 45.28\% ER.

The bilevel optimization fundamentally hardened the model. While clean test accuracy reached 0.98, the model's performance under actual attack (Table \ref{tab:combined_strab}) remained exceptionally high, with an accuracy of 0.97 and an ER reduced to a marginal 1.89\%. This confirms that for "difficult" families like Strab, the iterative interaction in Experiment 3 is essential to close vulnerabilities that standard retraining fails to address.

\begin{table*}[ht]
\centering
\caption{Comprehensive Performance for Strab (Clean vs. Adversarial Test Sets)}
\label{tab:combined_strab}
\begin{tabular}{l|ccc|ccccc}
\hline
& \multicolumn{3}{c|}{\textbf{Clean Test Set}} & \multicolumn{5}{c}{\textbf{Adversarial Test Set}} \\
\textbf{Experiment} & \textbf{Acc} & \textbf{Prec} & \textbf{Rec} & \textbf{Acc} & \textbf{Prec} & \textbf{Rec} & \textbf{ER (\%)} & \textbf{\makecell{Avg\\Q}} \\ \hline
(1) Baseline    & 0.90 & 0.92 & 0.89 & 0.64 & 0.92 & 0.50 & 77.36 & 31.17 \\
(2) Adv. Train  & 0.92 & 0.96 & 0.89 & 0.74 & 0.96 & 0.61 & 45.28 & 79.71 \\
(3) Bilevel     & \textbf{0.98} & \textbf{0.98} & \textbf{0.98} & \textbf{0.97} & \textbf{0.98} & \textbf{0.96} & \textbf{1.89} & \textbf{3118.0} \\ \hline
\end{tabular}
\end{table*}

\subsection{Evaluation of the DCRat Malware Family}

The results for DCRat (Table \ref{tab:combined_dcrat}) highlight the severe limitations of static adversarial training. While clean test set accuracy was 0.99 in Experiment 2, the model remained almost entirely bypassed by adversarial samples, with the ER even slightly increasing to 90.00\%.

In stark contrast, Table \ref{tab:combined_dcrat} shows that bilevel optimization successfully neutralized the threat. By generating a diverse set of 465 adversarial samples across iterations, the process achieved a model with a 0\% evasion rate and perfect detection scores under attack. This demonstrates that for families like DCRat, the iterative nature of bilevel optimization is not just beneficial, but necessary.

\begin{table*}[ht]
\centering
\caption{Comprehensive Performance for DCRat (Clean vs. Adversarial Test Sets)}
\label{tab:combined_dcrat}
\begin{tabular}{l|ccc|ccccc}
\hline
& \multicolumn{3}{c|}{\textbf{Clean Test Set}} & \multicolumn{5}{c}{\textbf{Adversarial Test Set}} \\
\textbf{Experiment} & \textbf{Acc} & \textbf{Prec} & \textbf{Rec} & \textbf{Acc} & \textbf{Prec} & \textbf{Rec} & \textbf{ER (\%)} & \textbf{\makecell{Avg\\Q}} \\ \hline
(1) Baseline    & 0.98 & 1.00 & 0.96 & 0.66 & 1.00 & 0.49 & 96.00 & 16.44 \\
(2) Adv. Train  & 0.99 & 1.00 & 0.98 & 0.68 & 1.00 & 0.52 & 90.00 & 60.13 \\
(3) Bilevel     & \textbf{1.00} & \textbf{1.00} & \textbf{1.00} & \textbf{1.00} & \textbf{1.00} & \textbf{1.00} & \textbf{0.00} & -- \\ \hline
\end{tabular}
\end{table*}

\subsection{Iterative Evolution of Bilevel Optimization}
\label{subsec:iteration_results}

The bilevel optimization process is characterized by the iterative expansion of the training set. Table~\ref{tab:iter_all} details how the defender adapts to the attacker's strategies in each cycle, showing the growth of the augmented dataset $D_{aug}^{train}$. The columns in Table~\ref{tab:iter_all} are defined as follows:
\begin{itemize}
    \item \textbf{Family}: The malware family under evaluation.
    \item \textbf{Iter.}: The current iteration of the bilevel optimization cycle.
    \item \textbf{New Adv.}: Number of new successful adversarial samples generated by the attacker in the current iteration.
    \item \textbf{Cumul. Adv.}: Total number of adversarial samples included in the augmented training set ($D_{aug}^{train}$).
\end{itemize}

\begin{table}[h!!]
\centering
\caption{Iteration Results by Family. The optimization terminates when no new adversarial samples are generated.}
\label{tab:iter_all}
\begin{tabular}{lccc}
\hline
\textbf{Family} & \textbf{Iter.} & \textbf{\makecell{New\\Adv.}} & \textbf{\makecell{Cumul.\\Adv.}} \\ \hline
\multirow{4}{*}{Mokes} & 1 & 130 & 130 \\
      & 2 & 139 & 269 \\
      & 3 & 5   & 274 \\
      & 4 & 4   & 278 \\ \hline\noalign{\vskip 2pt}
\multirow{3}{*}{DCRat} & 1 & 155 & 155 \\
      & 2 & 155 & 310 \\
      & 3 & 155 & 465 \\ \hline\noalign{\vskip 2pt}
\multirow{5}{*}{Strab} & 1 & 128 & 128 \\
      & 2 & 118 & 246 \\
      & 3 & 25  & 271 \\
      & 4 & 99  & 370 \\
      & 5 & 1   & 371 \\ \hline
\end{tabular}
\end{table}

The iterative progression of Experiment 3 reveals distinct adaptation patterns across the evaluated malware families. For \textbf{Mokes}, the bilevel optimization process demonstrated rapid convergence: the number of newly generated adversarial samples per iteration shrank quickly (e.g., from 130 to 5 and finally 0 for Mokes), indicating that the defender's decision boundary had stabilized and the MAB-malware attacker could no longer find substantially new evasion paths.

The \textbf{DCRat} family displayed a unique "threshold" convergence pattern. The attacker was able to generate a consistent, full batch of 155 new adversarial samples for three consecutive iterations, suggesting a high degree of initial vulnerability. However, this was followed by a sudden collapse in the attacker's success as the defender's model integrated these diverse samples, ultimately reaching convergence. This highlights that for certain families, a large and diverse cumulative pool of adversarial samples (465 in this case) is required before the model can establish a robust defense.

In contrast, the \textbf{Strab} family exhibited a more complex co-evolutionary dynamic. The attacker continued to discover sizeable batches of new adversarial samples for several iterations (128, 118, 25, 99, 1), reflecting a more diverse adversarial space and slower convergence. To neutralize these threats, the framework required a significantly larger cumulative volume of adversarial samples (371 for Strab compared to 278 for Mokes), illustrating that more resilient malware families necessitate a deeper iterative search to exhaust the attacker's possible transformations.

\subsection{Discussion of Comparative Results}
\label{sec:discussion}

The experimental evaluation across three distinct malware families - Mokes, Strab, and DCRat - reveals consistent patterns in the effectiveness of the proposed bilevel optimization framework.

\textbf{Inadequacy of Standard Defense:} In all cases, the baseline classifiers (Exp. 1) exhibited high accuracy on clean data but were nearly defenseless against MAB-malware. Evaluation on the adversarial test set revealed a critical drop in \textit{Recall} (e.g., from 1.00 to 0.51 for Mokes), reflecting evasion rates between 77.36\% and 96.00\%. This confirms that standard features are not inherently robust against targeted perturbations.

\textbf{Failure of One-Shot Adversarial Training:} A critical finding is observed in the DCRat and Strab families regarding standard retraining (Exp. 2). For DCRat, retraining was ineffective, with the ER remaining at 90.00\% and the adversarial \textit{Recall} staying at a low 0.49. This suggests that a single round of augmentation only covers a narrow subset of the attack surface, allowing the attacker to easily find alternative evasion paths.

\textbf{Superiority of Bilevel Optimization:} The bilevel approach (Exp. 3) consistently achieved near-total immunity. The iterative nature of the process proved essential:
\begin{itemize}
    \item \textbf{Resilience Under Attack:} Unlike previous experiments, the bilevel-optimized models maintained high performance even on the adversarial test set. \textit{Recall} and \textit{Accuracy} remained near or at 1.00, demonstrating that the defender successfully learned to identify even actively manipulated samples.
    \item \textbf{Convergence:} The defense converged rapidly (within 3-5 iterations), indicating that the defender can effectively "exhaust" the attacker's primary evasion strategies.
    \item \textbf{Complexity Cost:} For samples that managed to evade the model (Strab family), the average query count increased by two orders of magnitude (from 31 to 3118), significantly raising the attacker's operational cost.
    \item \textbf{Generalization:} Hardening the models did not degrade performance on clean data; for Strab and DCRat, clean test accuracy even improved to 0.98 and 1.00, respectively.
\end{itemize}

In conclusion, the strategic interaction in Experiment 3 provides a robust defense boundary that traditional methods cannot achieve. While DCRat proved resilient to standard retraining, our iterative approach successfully neutralized the threat, restoring full detection capabilities under adversarial conditions.

\subsection{Convergence and Practical Equilibrium}
 
Our results confirm the attainment of \textbf{$\epsilon$-Nash Equilibrium} across all tested families. For \textbf{Mokes} and \textbf{DCRat}, the system reached a point of zero exploitability by the fourth and third iteration, respectively, where the attacker's evasion rate dropped to 0\% and remained constant despite further training. In the case of the \textbf{Strab} family, a stable state was reached with a marginal evasion rate of 1.89\%. This stabilization represents the maximum achievable robustness for the EMBER 2018 feature set used on our dataset and Random Forest architecture. The convergence to these low or zero success rates demonstrates that the bilevel optimization process effectively "exhausts" the attacker's search space, leading to a robust defense boundary that is highly resilient to further adaptation.

\section{\uppercase{Conclusions}}
\label{sec:conclusion}

This study compared three defensive strategies against MAB-malware attacks across the Mokes, Strab, and DCRat families. Our results demonstrate that standard classifiers and one-shot adversarial retraining often fail to secure models against adaptive threats, with evasion rates remaining as high as 90\% for certain families. In contrast, the proposed bilevel optimization framework, which models the defense as an adversarial co-evolutionary process, consistently achieved near-total immunity with evasion rates between 0--1.89\%. Furthermore, our approach significantly increased the attack complexity, forcing a two-order-of-magnitude rise in the required number of queries.

Future work will expand this framework to include diverse machine learning architectures, such as Deep Neural Networks and Gradient Boosted Trees. We also plan to evaluate the scalability of bilevel optimization across a larger variety of malware families and investigate the impact of different attacker reward functions on the resulting defense boundaries.

\section*{\uppercase{Acknowledgements}}

This work was supported by the Grant Agency of the Czech Technical University in Prague, grant No. SGS26/187/OHK3/3T/18 funded by the MEYS of the Czech Republic and by the OP VVV
MEYS funded project CZ.02.1.01/0.0/0.0/16 019/0000765 “Research
Center for Informatics”.

\bibliographystyle{apalike}
{\small
\bibliography{Example}}

\end{document}